# Exploring aperiodic, complexity and entropic brain changes during non-ordinary states of consciousness

Victor Oswald[1,2], Karim Jerbi[2], Corine Sombrun[3], Hamza Abdelhedi[2], Annen Jitka[4,5,6], Charlotte Martial[4,5], Audrey Vanhaudenhuyse[1,7*], Olivia Gosseries[1,4,5*]

1. Conscious Care Lab, GIGA Consciousness, GIGA Institute, University of Liège, Liège, Belgium
2. Cognitive & Computational Neuroscience Lab, Psychology Department, University of Montréal, Montréal, Canada
3. TranseScience Research Institute, Paris, France
4. Coma Science Group, GIGA Consciousness, GIGA Institute, University of Liège, Liège, Belgium
5. NeuroRehab & Consciousness Clinic, Neurology Department, University Hospital of Liège, Liège, Belgium
6. Department of Data Analysis, University of Ghent, Ghent, Belgium
7. Interdisciplinary Algology Center, University Hospital of Liège, Liège, Belgium

*Contributed equally

Corresponding authors:

Victor Oswald– Conscious Care Lab, GIGA Consciousness, GIGA Institute, University of Liège, avenue de l'Hôpital 1, B34, B 4000 Liège, victor.oswald@umontreal.ca

Olivia Gosseries, Coma Science Group, GIGA Consciousness, GIGA Institute, University of Liege, avenue de l'Hôpital 1, B34, B - 4000 Liège, ogosseries@uliege.be

Audrey Vanhaudenhuyse, Algology Interdisciplinary Center CHU Liège - B35, and Conscious Care Lab, GIGA Consciousness, GIGA Institute, University of Liege, avenue de l'Hôpital 1, B34, B - 4000 Liège avanhaudenhuyse@chuliege.be

# Abstract:

Non-ordinary states of consciousness (NOC) provide an opportunity to experience deeply intense, unique, personal, and perceptually rich subjective states. Still, the neural mechanisms underlying these experiences remain poorly understood. This study aims to investigate the brain mechanisms associated with NOC, particularly those involving complex subjective experiences.

Twenty-seven participants, highly trained in Auto-Induced Cognitive Trance (AICT)—a self-induced, pharmacologically substance-free NOC—were recruited for this study. Participants underwent high-density EEG recordings during rest and AICT conditions. The aperiodic component (1/f) of the power spectral density, along with Lempel-Ziv complexity and sample entropy, was extracted from 5 minutes of clean signal for each condition. A multi-feature machine learning framework was employed to classify the two conditions and to identify the most important features through source-localized regions of interest. A subsequent analysis was performed to compare the EEG metrics between rest and AICT and to examine the influence of baseline levels on the changes observed between baseline and AICT, with a focus on investigating interindividual effects.

Whole-brain machine learning classification revealed significant changes in the aperiodic spectral component exponent (DA = 65%; $p < 0.05$), brain complexity (DA = 68%; $p < 0.01$), and entropy (DA = 70%; $p < 0.01$) during AICT compared to rest conditions. The comparison of EEG metrics demonstrated the greatest discriminative power for the aperiodic spectral component, followed by sample entropy and Lempel-Ziv complexity. Our analysis identified multiple frontal cortical areas, the posterior cingulate cortex, and the left parietal cortex as being predominantly involved in the AICT condition. Notably, resting state neural activity in the frontal and parietal regions played a significant role, serving as predictors of the magnitude of change between Rest and AICT.

These findings demonstrate that AICT involves the activation of brain regions known to support rich subjective experiences and offer a mechanistic perspective on how such experiences can influence brain functioning. This understanding opens new avenues for potential clinical applications.

# Introduction

Non-ordinary states of consciousness (NOC) have recently fascinated scholars and scientists across disciplines, offering a unique window into the complex dynamics and potential plasticity of the human brain and promising clinical interventions [1]. NOC can be induced by a myriad of modes of induction, such as meditative practices [2], the intake of psychoactive substances (e.g., psychedelics) [3], hypnotic suggestion [4,5], specific cognitive interventions [6,7], or heightened bodily functions (e.g., controlled breathing; [8–10]). NOC encompass states that differ from everyday consciousness, which is typically characterized by 'ordinary phenomenology' associated with conscious reasoning, thinking, and planning. NOC are characterized by high levels of absorption [5,11,12], pronounced dissociation [2,4,13,14], and other nuanced cognitive shifts such as dereification and self-decentering [2,15–17].

Beyond their rich subjective experiences, NOC exhibit tangible modulations in physiological systems, notably the autonomic nervous system [18–22] and central nervous system [19,23–30]. A common approach to studying NOCs involves examining the physiological changes they induce, often reflected in dynamic shifts in brain activity and functionality. These cerebral changes are diverse and can be explored through various methods. Some changes are state-related [31,32], while others reflect long-term neuronal modifications resulting from consistent practice (i.e., trait-related) [33–35]. NOC-related states are often associated with increased neural complexity and entropy. This can arise from enhanced bottom-up sensory and interoceptive processing (e.g., pharmacological or sensory-driven states), but also from a relaxation of top-down predictive control and self-induced processes (e.g., meditation, hypnosis). Current models suggest that the interaction between reduced top-down constraints and increased bottom-up signaling together contributes to the observed NOC [24,29,36,37]. These neural changes may also be associated with change in the excitatory/inhibitory (E/I) ratio [38], reflected in a reduction of the aperiodic component of the power spectrum (commonly referred to as the 1/f slope) [39,40]. Conversely, there is often a reduction in higher-order top-down oscillatory activity, accompanied by a flattening of the brain's functional hierarchy [41,42]. This is thought to arise from a general decrease in neural inhibition, which contributes to a more disorganized yet flexible state of brain activity [41,43]. While this general framework provides a clearer understanding of brain activities during NOCs, the research landscape is marked by heterogeneous studies reporting divergent results, reflecting differences in studied NOC types, practiced techniques, individual baselines, and recording conditions.

One such NOC that has recently garnered attention is the Auto-Induced Cognitive Trance (AICT) [7,44,45]. AICT is characterized by its self-evoking nature, allowing individuals to enter a modified state devoid of exogenous substances or external stimuli. Typically, AICT involves intentional modulation of attentional focus, often inward, leading to experiences ranging from deep absorption to intense dissociation from the immediate environment [7,44–47]. Upon adequate training, most of the individuals attain autonomy, enabling them to self-induce (via movement, vocalization and will) and harness the AICT technique for diverse purposes. While preliminary studies have documented shifts in heightened phenomenology [7,44,47], modulations in brain functioning [29,48], and changes in autonomic nervous system activity [49], our recent work has also highlighted the key role of interindividual variability, suggesting a self-regulatory mechanism underlying AICT [49]. However, the neural mechanisms that sustain these subjective experiences during this NOC remain poorly understood.

In the last decade, research on consciousness has led to the development of methods to characterize the quantity and complexity of information carried by neural signals [50–52]. This complexity appears to reflect the level of information processed by the brain and may represent an underlying mechanism for the content and information accessible to consciousness [38,53,54]. This approach notably allows for distinguishing between patients with disorders of consciousness and under anesthesia [55,56]. These same methods have also been used to differentiate ordinary states of consciousness from NOC induced by meditation and psychedelics [35,41,57].

The aim of this study is to characterize the neural correlates of AICT by assessing brain activity and brain diversity. Specifically, we examine whether brain electrical activity during AICT differs from baseline states and how interindividual variability shapes these effects. Based on prior work linking altered states of consciousness to heightened signal diversity, we hypothesized that AICT would be associated with change in global brain activity, complexity and entropy.

# Methods

This study utilized a previously published dataset [7,20,44,48,49]. The initial sections of the methodology, including Participants and Training, Procedure, were adapted and reformulated based on prior work [20,49].

### Participants and training

The study included 27 adult participants (23 females; mean age: 45 years ± 13 years; range: 24–72 years), all native French speakers. Participants were highly experienced in entering the AICT state, with an average of 28 months of practice (± 39 months; range: 9–216 months). They underwent prior AICT training, which involved a sound-loop-based program designed to facilitate voluntary induction of the AICT state without reliance on sound or movement cues. During training, participants lay on the floor with their eyes closed, listening to sound loops developed by the TranceScience Research Institute (www.trancescience.org). These loops featured electronic binaural sounds (pure tones between 100Hz and 200Hz with beat rates below 10Hz) and voices. AICT could also be induced through various methods such as vocalizations (e.g., singing, protolanguage) or specific movements (e.g., stereotyped gestures). Participants progressively identified their preferred inducer (e.g., hand movements, vocalizations) and were encouraged to practice independently at home and without the help of the sound loop. Participants were fully informed about the study's objectives and provided written consent before participating. No incentives were offered for participation. The study was approved by the Ethics Committee of the Faculty of Medicine at the University of Liège (reference 2019/141) and adhered to relevant ethical guidelines and regulations.

### Procedure

Before the experiment, demographic data, including age and sex, were collected. To facilitate concurrent electroencephalography recordings, participants were trained in advance and instructed to remain motionless during their trance states. The experimental session comprised five conditions: Resting state (Rest); Ordinary conscious state with auditory stimulations (Auditory); Imagining a previous intense AICT without entering a trance (Imag); AICT and AICT with auditory stimulations (Auditory-AICT). The first three conditions were counterbalanced across participants, while the final two conditions were conducted sequentially to prevent after-effects of AICT from influencing the earlier tasks. During each condition, participants kept their eyes closed. In the Rest and Auditory conditions, participants allowed their thoughts to flow freely. In the AICT and Auditory-AICT conditions, participants used their preferred induction techniques to enter and maintain the AICT state. These techniques often involved body movements or vocalizations. Participants remained motionless once in a trance state; if the state faded, they were free to re-induce it through movement or vocalization, after which they returned to stillness. Recording durations were adjusted accordingly. Each condition lasted approximately 12 minutes. After each session, participants completed a free-recall task and self-report questionnaires, including whether they reached a trance state (yes/no) and if so, rate its intensity on a Likert scale ranging from 0 (no trance) to 10 (the most intense trance ever experienced) (Trance Intensity). Various classic phenomenological measures typically associated with NOC were also collected (see others works [7,44,58]).

This study focused specifically on two conditions: the baseline resting state (Rest) and AICT, excluding the Imagination and the two Auditory conditions due to their relevance to a separate project.

**Electroencephalography recordings & data preprocessing**

During both conditions, cerebral activity was recorded using a high-density EEG system with 256 electrodes (EGI, Electrical Geodesics, USA). The EEG data were pre-processed using Brainstorm software [59]. The pre-processing steps included re-referencing, the rejection of 84 neck and facial channels, notch filtering power line at 50 Hz and 100 Hz, and bandpass filtering between 0.5 and 60 Hz (Figure 1). The data were then downsampled from a sampling rate of 500Hz to 240 Hz. Artifact rejection was performed using Independent Component Analysis, and the resulting components were visually inspected to identify and reject any artifact-related components. For further analysis, windows of interest in the EEG signal were manually selected and concatenated, ensuring that 4 minutes of clean signal were retained for each condition. For the AICT condition, the time windows were selected from periods outside of the induction period.

Physiological data, including electrocardiogram (ECG), and respiration were recorded using the EGI polygraph input box. ECG and respiration data was previously published [20,49]

**Cortical source reconstruction**

The MNI ICBM152 brain template was utilized as the default anatomy T1-MRI and head surface for individual cortical level source reconstruction (Figure 1). To compute the head model for forward modeling of the magnetic field, an overlapping-sphere method with OpenMEEG was employed [60]. The weighted minimum norm imaging solution was computed with a dipolar orientation constraint to the cortex, a signal-to-noise ratio of 3 and a use depth weighting at 0.5. The noise covariance matrix was estimated for each participant and condition using only the diagonal elements, which correspond to the variance measured at each sensor computed from recording. The reconstructed source time series were initially generated on a 15,000-vertex individual brain tessellation [61,62].

**Feature Extraction**

To investigate state-dependent changes in EEG during AICT, we extracted aperiodic (1/f) components of the power spectral density along with measures of neural complexity and entropy. These features were selected for their relevance to NOC and for their capacity to reflect aspects of the temporal organization of brain activity. By focusing on these metrics, we aimed to capture summary indices of how neural signals are organized over time and to identify neural signatures associated with AICT.

<u>Aperiodic (1/f) components</u>

The power spectrum density (PSD) was calculated using a modified Welch periodogram technique with a 5-second Hamming window and 50% time-window overlap at each cortical source. To disentangle and parameterize the periodic and aperiodic features of the PSD spectrum, the Fitting Oscillations and One-Over-F (FOOOF/specparam) algorithm, developed by [63], was employed. The FOOF model was computed with between 1Hz-60Hz, with maximum peak at 3, detecting peak type with gaussian approach, proximity threshold at 2, and knee option for aperiodic mode selection. The 1/f model was subtracted from the full spectrum. Thus, we extracted 1/f exponent, which reflect power law of the power spectrum and has previously been shown to reflect changes in E/I balance [38,54]. The PSD model was computed for each vertex, and the specparam values were then averaged across 7,502 cortical vertices.

Complexity (Lempel-Ziv)

The Lempel-Ziv complexity (LZC) is a measure used to assess the regularity or complexity of a signal. It involves scanning symbolic sequences within the signal to identify new patterns and incrementing the complexity count whenever a new sequence is detected. Regular signals tend to have fewer distinct patterns, resulting in a lower LZC, while irregular signals exhibit a higher LZC due to a greater number of unique patterns. For complexity analysis, we utilized Neurokit2, a Python package known for its advanced biosignal processing capabilities [64]. To begin, we applied a window size of 1 minute to the signal. Within each window, we symbolized the signal using a median split technique. This involved dividing the window into segments based on the median value and assigning symbols to each segment. Once the signal was symbolized, we computed the LZC using a delay of 1 and a dimension of 2, as the default parameters. The complexity calculation was performed on each vertex, resulting in a total of 7502 vertices, for each 4-minute segment of the signal. Finally, we determined the average complexity across the windows, providing an overall measure of complexity for the analyzed signal.

Sample Entropy

Sample entropy is a widely used measure for assessing the complexity of physiological time series signals [65]. It quantifies the likelihood that two vectors, close to each other in m dimensions, will remain close in the next m+1 component. We employed Neurokit2 [62] to perform sample entropy analysis. We applied a window size of 1 minute to the EEG signal. We computed the sample entropy with a delay of 1, a dimension of 2, and a tolerance set to 0.2 times the standard deviation of the signal, which are the default parameters. The sample entropy calculation was performed on each vertex, resulting in a total of 7502 vertices, for each 4-minute segment of the signal. By calculating the average entropy across the windows, we aim to capture the overall complexity of the EEG signal in both conditions. This approach provides insights into the variability and information content of the signal, allowing us to investigate potential changes in complexity associated with the AICT state.

Parcellations

To simplify comparisons across regions of interest (ROIs) and streamline the computational process, all EEG features computed at the source level were downsampled to a 300-ROI template based on the Yeo resting-state atlas, which provides a distributed parcellation of the entire cerebral cortex [66]. This downsampling approach standardizes the data across different ROIs and reduces the computational workload, making the analysis more efficient and interpretable. Ultimately, it enables easier comparisons and extraction of meaningful insights from the EEG data.

**Supervised machine learning analysis**

Multi-feature classification across brain space

Instead of exploring a single ROI model, we decided to implement a more comprehensive multi-feature classifier that incorporates all ROIs as features to build a single model for each EEG feature (Figure 1). To assess the individual contributions of each ROI, we employed a random forest algorithm [67], which estimates feature importance by averaging the relative ranks of each feature across the decision trees within the forest. To measure the variability in feature importance, we repeated the training process for n=100 times. Within each iteration, we utilized grid search with 5-fold cross-validation as part of a nested cross-validation approach, wherein a different subject was excluded for testing in each iteration. The overall model score was determined using the test samples from a 5-fold cross-validation procedure. By employing this approach, we were able to gain valuable insights into the significance of each ROI in influencing the overall performance of the classifier. The combination of random forest and cross-validation techniques allowed us to obtain robust and reliable feature importance estimates for our multi-feature classifier.

Multi-feature classification across brain space and features extractions

To systematically compare the contribution of various EEG metrics to the decoding of conditions between *Rest* and *AICT*, we performed a multi-feature classification analysis across brain regions by integrating all EEG metrics. Specifically, we combined 300 ROIs associated with the 1/f component, 300 ROIs with LCZ, and 300 ROIs with sample entropy, resulting in a dataset comprising 900 features. Using the previous established classification framework, we evaluated the combined dataset to determine the feature importance of each ROI across the 900 features. This approach allowed us to rank the ROIs based on their importance, providing a comprehensive hierarchy of ROI contributions across all EEG metrics investigated.

Statistical significance of the classification

To evaluate the statistical significance of the classification results, we employed a binomial test to determine whether the observed accuracy exceeded chance levels (50% for binary classification). The mean classification accuracy across all iterations was calculated to represent the average proportion of correct predictions. Based on the total number of predictions, the estimated number of correct predictions was derived by multiplying the mean accuracy by the total number of true labels. Under the null hypothesis, the probability of a correct prediction by chance was set at 0.5. Using the cumulative distribution function of the binomial distribution, we calculated a global p-value, representing the probability of achieving the observed number of correct predictions or more by chance. A p-value below the conventional threshold (e.g., 0.05) indicates that the model's performance is significantly better than random chance, providing robust evidence for the validity of the classification results.

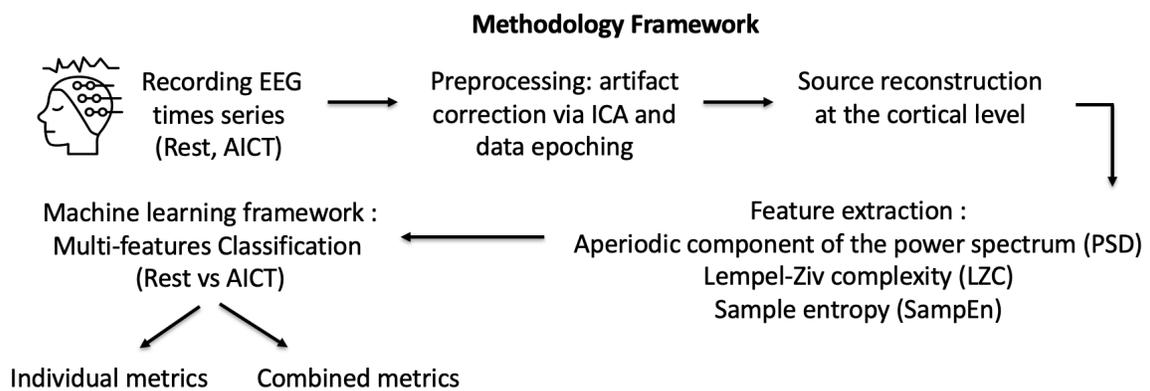

*Figure 1. Methodology framework*

**Inter-individual variability analysis**

To investigate inter-individual differences between the baseline state (Rest) and AICT, we applied a linear mixed model with two measurement points (i.e., Rest and AICT) for each ROI feature. Additionally, we examined the relationship between the intercept and the slope across the two time points. The linear coefficient between these parameters was computed using the following formula:

$$Coef(int, slope) = \frac{cov\,(int, slope)}{sd(int) * sd(slope)}$$

This coefficient was calculated for each feature across all participants. To assess the statistical significance of these coefficients, we generated a null distribution using a bootstrap method with 400 repetitions to derive confidence intervals for each coefficient. Finally, we applied this analysis to all ROIs, reporting only the coefficients that fell outside their corresponding confidence intervals, thereby identifying statistically significant results.

# Results

All participants reported successfully reaching a trance state during the AICT condition, with a mean self-reported intensity of 6.72 (SD = 1.77; range = 3–10) [7,20,44,48,49].

**Aperiodic change between Rest and AICT**

Figure 1a presents the mean full power spectrum for the Rest and AICT conditions across participants. The aperiodic component of the power spectrum (Figure 2b) shows a decrease in the 1/f exponent during AICT, visible as a flattening of the aperiodic slope compared to rest. Figure 2c presents the periodic component extracted from the same model. Individual and group mean values of the 1/f exponent and offset are shown in Figures 2d and 2e, respectively. AICT is associated with a reduced 1/f exponent, while no significant change is observed in the PSD offset.

Machine learning classification between rest and AICT across the entire brain space achieved notable performance for the 1/f exponent, with an accuracy of 65% (SD = 0.1; p

= 0.02) (Figure 4). The spatial effects between conditions, assessed using Cohen's d, are shown in Figure 3a, with the largest effect observed (d>0.8) in the left frontal regions, where the AICT condition exhibited a decrease in the 1/f aperiodic exponent. These results align with the feature importance map extracted from the random forest classification between AICT and rest (Figure 5a), which identified the left dorsal frontal region as one of the most contributing ROIs. Other brain regions also exhibited congruent changes and heightened feature importance, including the bilateral dorsomedial and cingulate frontal areas, as well as the bilateral posterior cingulate and the left temporo-occipital junction.

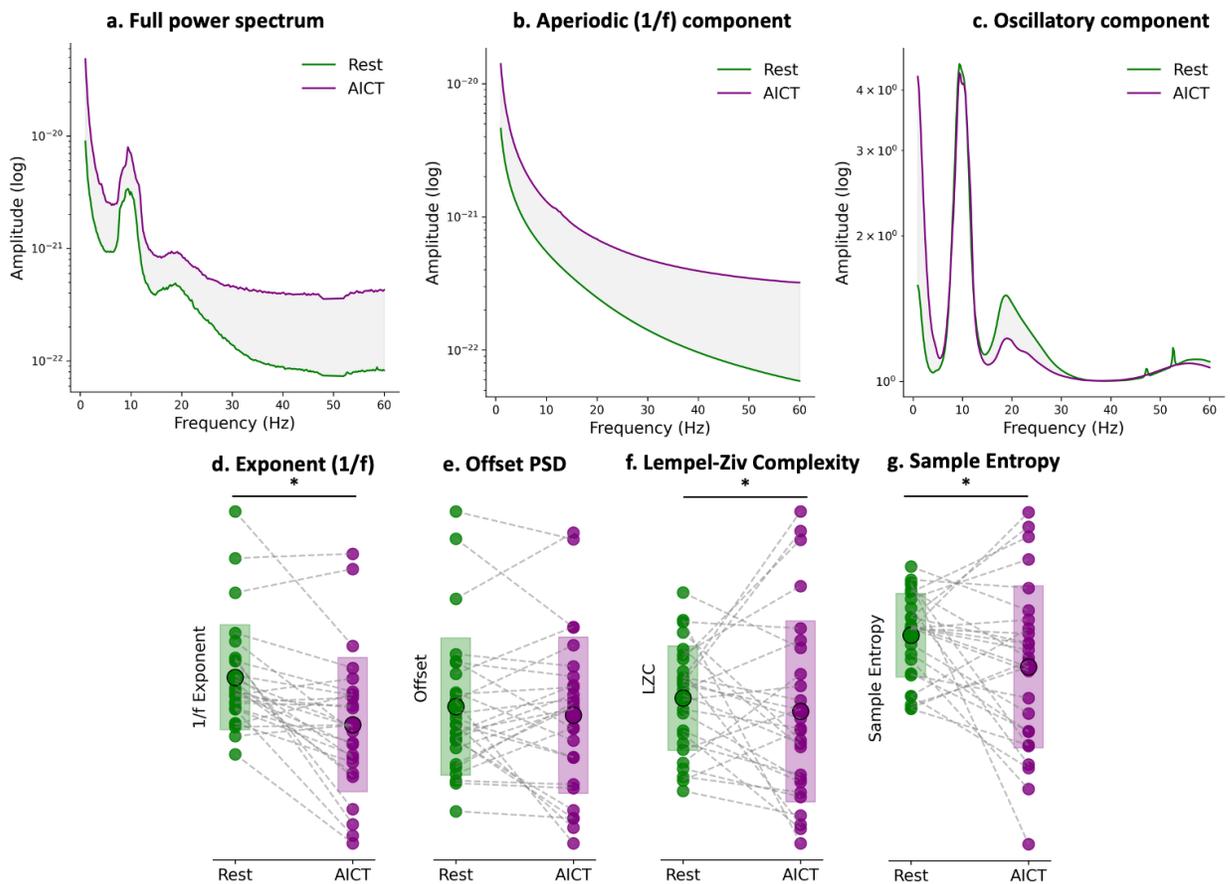

*Figure 2: Power spectral density (PSD) across all participants for each condition, Rest (green) and Auto-induced cognitive trance (AICT, purple). Mean across all participants (n=27) full power spectrum (Figure 2a); aperiodic (1/f) component modeled using the FOOOF model (Figure 2b); periodic component (Figure 2c); individual and group mean values for the 1/f exponent (Figure 2d), PSD offset (Figure 2e), Lempel-Ziv Complexity (Figure 2f) and Sample Entropy (Figure 2g).*

**Brain complexity changes between Rest and AICT**

Overall, LZC metrics revealed a substantial decrease in neural complexity (Figure 2f). This effect, however, showed spatial heterogeneity: during AICT, complexity decreased in parietal and occipital regions, while bilateral frontal and temporal areas exhibited increased complexity compared to Rest. The most pronounced reduction was observed in the left parietal cortex (d = 0.6; Figure 3b).

Machine learning analysis achieved a decoding accuracy of 68% (SD = 0.11; p = 0.009) (Figure 4a). The spatial distribution of feature importances highlighted a predominant effect in the left parietal regions, followed by bilateral frontal regions (Figure 5b). These regions appear to target distinct brain areas compared to those associated with the 1/f exponent, suggesting a differential involvement of neural functioning in these metrics.

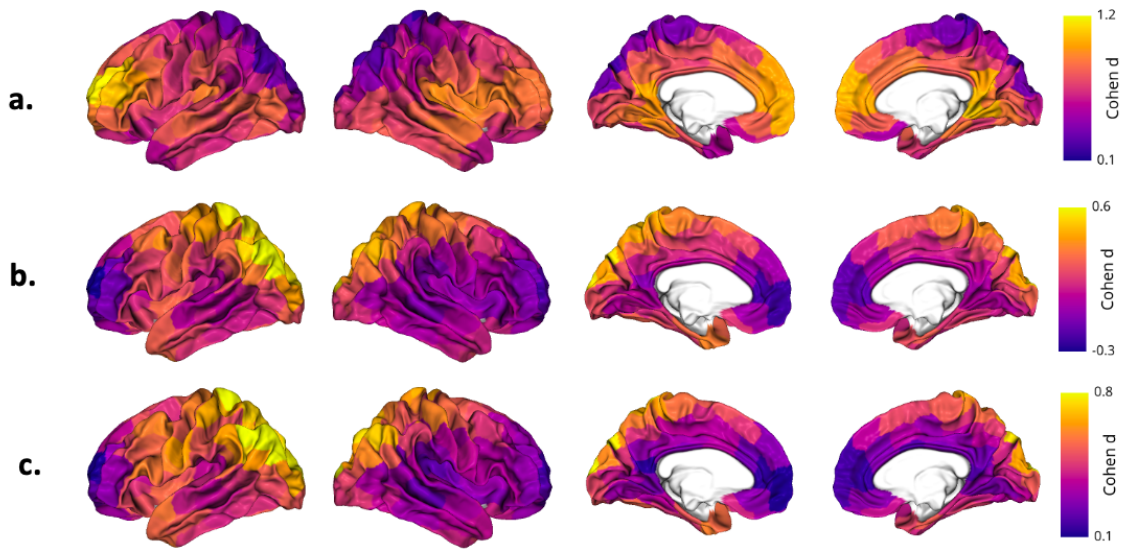

*Figure 3: d' Cohen effect brain map between Rest and Auto-induced cognitive trance (AICT). a. 1/f Aperiodic exponent b. Lempel-ziv complexity and c. Sample entropy*

**Brain entropy changes between Rest and AICT**

We observed a prominent decrease in brain entropy across the entire brain during AICT (Figure 2g), with the strongest effects (d>0.8) localized in the left dorsal regions, specifically the left parietal and left occipital-parietal areas (Figure 3c).

Machine learning analysis demonstrated a decoding accuracy of 70% (SD = 0.1; p = 0.009) (Figure 4a). The spatial distribution of feature importances identified a predominant effect in the left parietal and occipital regions. These entropic changes during AICT compared to Rest closely align with findings derived from the brain complexity approach.

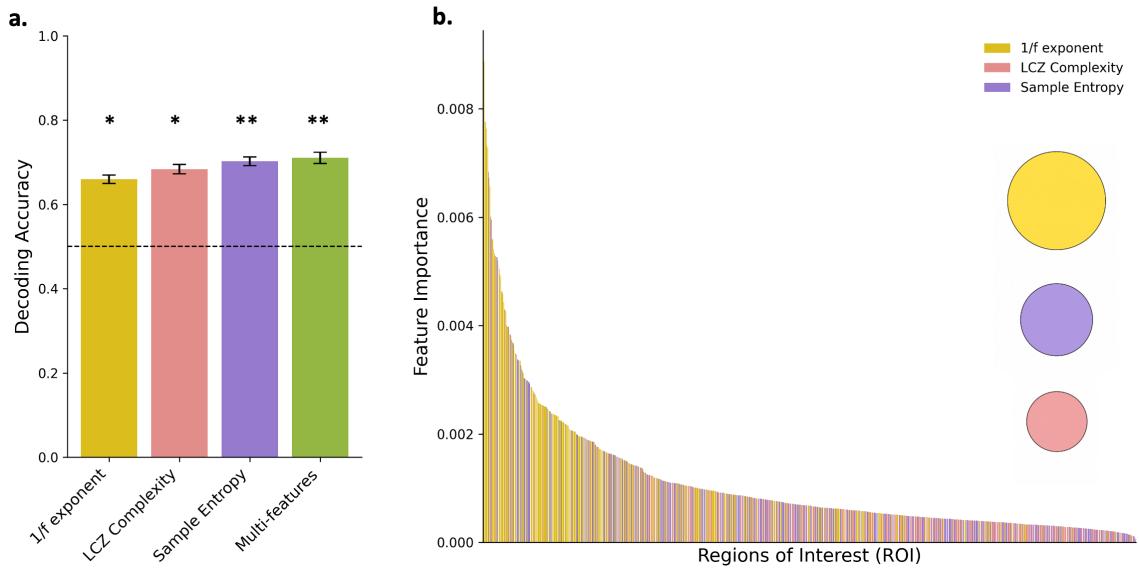

*Figure 4: Multi-feature Machine Learning Classification between Rest and Auto-Induced Cognitive Trance (AICT). **a.** Decoding accuracy cross validated framework using random forest regressor to classify (rest vs. AICT) for 1/f Aperiodic exponent, Lempel-Ziv complexity, sample entropy and all multi features with the standard deviation. All features show significance with * <.05, ** <.01. **b.** Multi-feature importance is illustrated after multi-feature across all regions of interest (300) and 3 metrics. Circle size represents the proportion that each metric contributes across the entire multi-feature classification.*

**Multi features analysis and features importance across all EEG metrics**

The multi-feature classification incorporated three EEG metrics across 300 regions of interest, resulting in a total of 900 features. The classification process was repeated 100 times, yielding a classification performance of 71% (SD = 0.13, p = 0.001). Feature importance values were extracted during each iteration, averaged, and ranked in descending order, as presented in Figure 4b. Notably, ROIs associated with the 1/f aperiodic component exhibited the highest feature importance, followed by ROIs linked to sample entropy and Lempel-Ziv complexity. These findings align with Cohen's *d* effect size values shown in Figure 3, reinforcing the consistency of the observed spatial patterns across metrics.

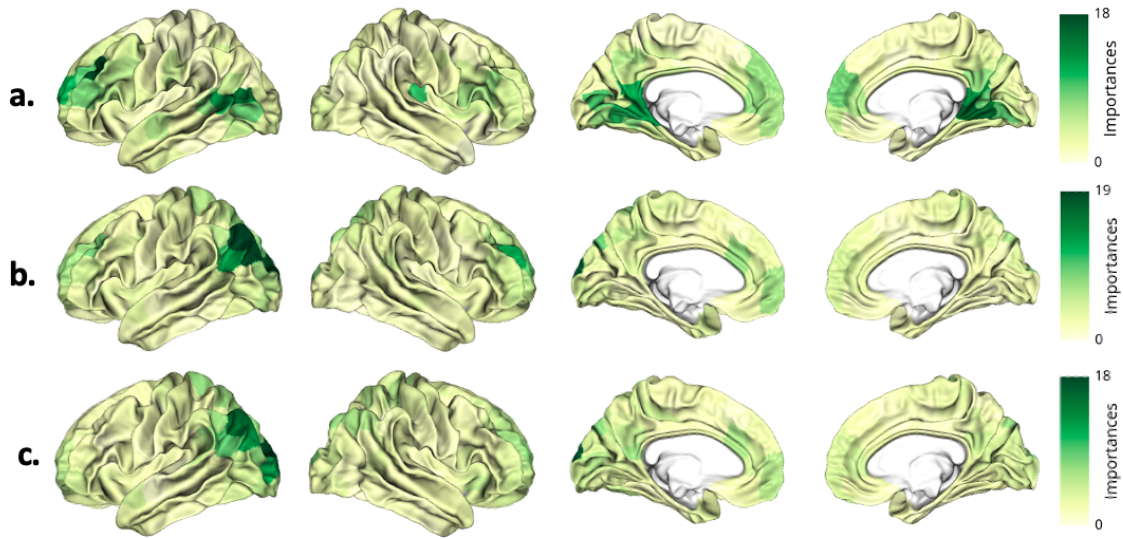

*Figure 5: Features importance extracted from random forest multi-feature classification across all brain space between Rest and Auto-induced cognitive trance (AICT) for a. 1/f Aperiodic exponent b. Lempel-ziv complexity and c. sample entropy.*

**Interindividual variability in brain change**

Investigation of interindividual variability in the relationship between slope and intercept from the linear model revealed ROIs with divergent trajectories across individuals and across metrics used. As illustrated in Figure 6 (and S1), the significant regions indicate that these effects are driven by certain participants more than others (-1SD vs +1SD). Notably, the AICT state appears to drastically reduce the interindividual differences observed during the resting state. For the aperiodic (1/f) component, significant ROIs were identified in frontal regions, as well as in the left caudal sensorimotor area, the left inferior parietal lobule, and the left superior temporal gyrus. The strongest coefficients were observed in bilateral frontal regions. For brain complexity, significant regions were found in bilateral occipital and parietal areas, with the strongest coefficients lateralized to the right. For brain entropy, a large number of significant regions were identified, including bilateral dorsal, parietal, occipital, and sensorimotor areas, as well as the anterior temporal poles and the left inferior frontal regions. The strongest coefficients were lateralized to the left, spanning a network of frontotemporal and parieto-occipital regions.

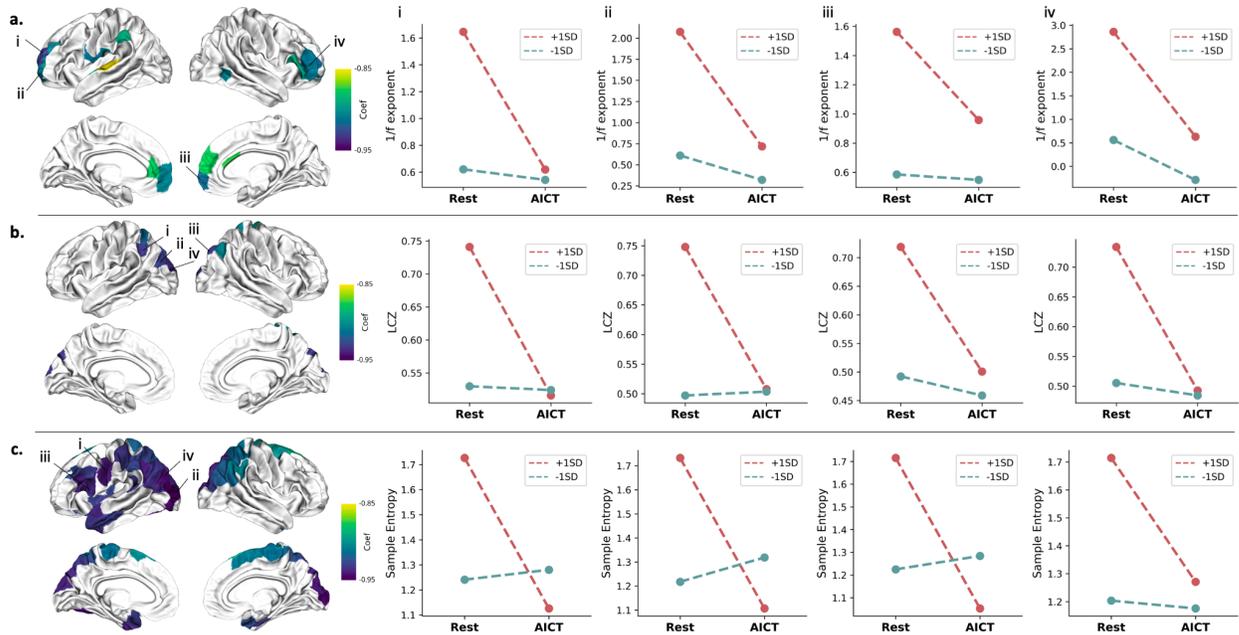

*Figure 6: Interindividual Variability in the Relationship Between Slope and Intercept from the Linear Model. Panels represent variability for (a) 1/f aperiodic exponent, (b) Lempel-Ziv complexity, and (c) sample entropy. The left panel highlights brain ROIs with coefficients that fall significantly outside the confidence interval (uncorrected). The right panel depicts plots of the four highest coefficients, illustrated with the mean ± 1 standard deviation for each condition. Values higher during Rest are shown in red, while those lower during Rest are shown in blue.*

# Discussion

Summary

This study shows distinct EEG changes during AICT compared to Rest. We observed a decrease in spectral power 1/f slope, Lempel–Ziv complexity, and sample entropy, reflecting neural modulation in the AICT state. The aperiodic slope flattening was particularly marked in the left frontal cortex, while Lempel–Ziv complexity, and sample entropy, revealed most size effect in left parietal regions. Inter-individual differences, and particularly the variability observed at rest, were linked to the magnitude of state-related changes across individuals.

Aperiodic (1/f) component changes in AICT

The change in the aperiodic (1/f) component between conditions reveals a significant decrease of PSD slope across the brain during AICT, with the most notable changes observed in the left frontal regions, left dorsolateral prefrontal cortex (DLPFC), bilateral medial frontal regions and posterior cingulate cortex (PCC), including the temporo-parietal junction (TPJ). Cohen's d effect size and feature importance scores converge on this set of regions, highlighting their critical role in the observed dynamics. The aperiodic component (1/f) is hypothesized to reflect the balance between excitation and inhibition within a neural network. A lower 1/f exponent reflect increased cortical excitability or arousal, depending on the context [54,63,68,69]. Thus, the AICT state appears to be characterized by global heightened cortical excitability, particularly in the frontal regions, which may be associated with enhanced cortical information integration [70].

The strongest effect is observed in the DLPFC, known as a crucial hub for accessing conscious information, as described by the Global Workspace Theory [71–73]. Indeed, other studies have shown that during AICT states, more information is represented in the individual's consciousness, as illustrated by an increased intensity of thoughts, emotions, and sensory perceptions [7,44]. Change in cortical excitability were also observed in the rostro-lateral prefrontal cortex (RLPFC), a key hub involved in switching attention between internal and external stimuli (e.g., between self-related processing and task- or goal-oriented processing) [74], which may underlie other phenomenal features associated with AICT states. In other NOC, such as long-term mediators, functional coupling between RLPFC and PCC was associated to better subjective attention [75].

Additionally, excitation in the PCC, a core region of the Default Mode Network (DMN), was found to increase his cortical excitability during AICT states. The PCC is central to introspective thinking, autobiographical memory retrieval, and disconnection from the external environment, all of which are strongly stimulated during AICT experiences [7,44]. The rostro-medial prefrontal cortex (RMPFC) was also found to be increased functionally through its cortical excitability, the RMPFC is involved in subjective feelings, autobiographical narrative, and it is connected to PCC through the DMN [76].

These frontal regions and the PCC belong to distinct and often antagonistic networks. The DLPFC and RLPFC, key hubs of the Central Executive Network (CEN), are typically active during externally focused, goal-oriented tasks. The CEN is also strongly associated with conscious access, enabling the selection and maintenance of information in consciousness. In contrast, PCC and RMPFC, core nodes of the DMN, are predominantly active during rest or internally oriented processes such as mind-wandering, autobiographical memory, and self-referential thinking [77–79]. Traditionally viewed as functionally segregated, these two networks have been found to co-activate during specific states involving intentional internal focus, such as meditation or active introspection[80]. This presents a paradox: although the CEN is typically associated with externally directed cognition, its activation in these contexts suggests that internally guided mental states also require active executive control—such as sustaining attention, inhibiting distractions, and monitoring internal content. This functional interaction between the CEN and DMN may facilitate perceptual decoupling and attentional redirection toward internal experience, both of which are phenomenal hallmarks of AICT states. Such CEN–DMN synchronization challenges the classical dichotomy between "task-positive" and "task-negative" networks, pointing instead toward a dynamic and cooperative model of conscious processing. This interaction could be further validated through future investigations of functional and dynamic connectivity between these two networks during AICT states. Consistent with this view, a previous study reported that shamanic practitioners in trance exhibited increased functional centrality of the PCC and stronger coupling between DMN hubs and executive control regions. These findings support the idea that altered states of consciousness engage both networks simultaneously, facilitating absorption and perceptual decoupling from external input [81].

Our findings target the same functional regions as those involved in other NOC, such as meditation and hypnosis. We report an increase in activity within DMN-related regions, including the PCC and the RMPFC, as well as other frontal regions associated with the CEN, such as the RLPFC and DLPFC. Interestingly, several studies on meditation and hypnosis have typically reported a decrease in DMN activity during these states [25,82–85], which is often associated with a reduction in mind-wandering [86]. In contrast, our findings reveal the opposite effect, suggesting that this difference is linked to the increased amplification of subjective experiences during the AICT state. DMN activity is not only associated with internally directed states, such as mind-wandering, but is also known for linking concepts that support introspection, imagination, creativity, and divergent thinking [87,88]. Therefore, the observed increase in DMN-related activity may reflect a deeply absorptive internal state, characterized by a rich and vivid inner subjective experience.

Despite some heterogeneity, meditation studies typically report increased activity in CEN regions as measured by fMRI, along with a decrease in the 1/f exponent during meditation [57], which has been linked to enhanced cognitive control [83,89]. In contrast, hypnosis is often associated with a decrease in dorsal anterior cingulate cortex activity, reflecting appraisal and expression of fear and pain, as well as a sense of personal agency [90,91]. We propose that in AICT, the frontal regions associated with the DMN play a central role in shaping the suggestive experience itself, while other frontal regions, such as the RLPFC, contribute to the conscious access and integration of this experience. Notably, the RLPFC is recognized for its role in the gating hypothesis [74], facilitating the incorporation of subjective experiences into conscious awareness. AICT is not only characterized by a rich inner experience but also requires the ability to filter distractions and maintain focus throughout the state. Additionally, it involves processing the emotional content experienced during the state [7]. These functions (integrating subjective experiences, sustaining focus, and processing emotional content) are strongly supported and specialized by the frontal regions highlighted in our results [76].

The phenomenon of experiencing one's subjective reality in an intensified and rich manner is a defining feature of the AICT state. This mechanism may serve as a crucial process for fostering deep engagement with inner experiences, offering valuable opportunities for self-reflection and insight when integrated into psychotherapeutic approaches [92]. Such enhanced engagement with subjective experiences holds transformative potential, particularly in therapeutic contexts that benefit from deep introspection and the integration of personal narratives [93,94]

Complexity and entropy change in AICT

The observed decrease in brain complexity and entropy, particularly in the left parietal cortex and left TPJ during AICT are congruent while brain complexity reports moderately increases in the frontal brain regions. Complexity and entropy metrics are conceptually interrelated, which accounts for their overlapping spatial patterns. However, our analysis also identifies a significant increase in the fronto-temporal regions, specifically on the complexity (LCZ) metric, suggesting a distinct functional specificity in these areas. The reduction in complex or entropic brain activity is associated with decreased integration of cortical inputs and less comprehensive information processing in the affected region [96,97]. A previous study reported similar results in shamanic trance, showing a decrease in gamma-band Lempel–Ziv complexity (LZC) compared to controls, and this decrease was correlated with increased feelings of insightfulness [95]. These functional changes may be linked to the state induced by AICT. Indeed, the left parietal cortex is notably involved in working memory, sequential and analytical processing (e.g., mathematical reasoning), as well as the execution of precise motor sequences [98–102]. These functions are particularly inhibited or absent during the AICT state, which promotes internalization and a loss of both mental and motor agency [7,44]. Additionally, the parietal regions belong to the CEN, which is often antagonistic to the DMN, which appears to be particularly dominant during this AICT, further supporting this functional shift.

The increase in brain complexity observed in the frontal and temporal regions indicate greater neuronal information exchange in these areas. This aligns with the previously discussed 1/f changes in frontal regions, which are associated with the processing of conscious information. Moreover, in a previous study conducted on a single participant, our team used a TMS–EEG paradigm to compare neuronal responses to TMS during rest and during AICT. The results pointed to increased TMS-evoked responses when stimulating frontal regions and decreased responses when stimulating parietal regions during AICT compared to rest. This pattern aligns with the increase in frontal complexity and decrease in parietal complexity reported here [29].

In the context of AICT, a decrease in neural entropy in the left TPJ may reflect a more stable and constrained mode of processing, aligning with phenomenological reports of diminished self-other boundaries, altered body awareness, and a sense of absorption. The left TPJ is critically involved in self-location, bodily ownership, and perspective-taking [103,104] and plays a key role in integrating multisensory information to maintain a coherent sense of self in space [105]. Reduced signal complexity in this region during AICT could thus reflect a functional decoupling from external sensory inputs, corresponding to the subjective immersion and disembodiment often described in AICT state.

<u>Convergent and Divergent EEG Markers of AICT</u>

Our results show that the aperiodic 1/f slope, Lempel–Ziv complexity (LZC), and sample entropy are interrelated yet capture distinct aspects of neural activity. These three metrics reveal both convergent and divergent patterns, making them complementary in the study of NOC. For instance, regions showing stronger decreases in the 1/f slope overlapped with those showing increases in LZC (e.g. frontal regions), while decreases in complexity often coincided with decreases in entropy—an expected finding given the known inverse relationship between the 1/f slope and signal complexity, as well as the partial correlation between LZC and entropy. Importantly, our multi-metric analyses demonstrated that combining all three measures increased discriminative power compared to each metric alone. Moreover, feature-importance analyses revealed that each metric emphasized distinct spatial regions, suggesting that despite their interdependence, they each capture unique aspects of cerebral dynamics.

<u>Inter-individual effects</u>

Interindividual effects reveal distinct brain regions that are highly sensitive to baseline variability (i.e., spontaneous brain activity). For instance, among the most significantly modulated regions—such as those associated with changes in 1/f dynamics—the frontal regions appear particularly responsive to interindividual variability between the baseline and AICT states. In contrast, posterior regions, notably the PCC, show relatively stable activation patterns across individuals. This pattern suggests that while some regions are consistently recruited during AICT regardless of individual differences, other areas—especially in the frontal cortex—adjust more flexibly as a function of the baseline state.

These findings may reflect individual differences in trait responsiveness to absorption and suggestion, concepts that have been extensively studied in the context of hypnotisability. High-hypnotisable individuals typically exhibit greater functional modulation in frontal regions during NOC, which has been interpreted as a marker of their enhanced top-down control and capacity for focused attention [106,107]. Our results are consistent with this framework, suggesting that variability in AICT-related frontal engagement may similarly reflect a spectrum of susceptibility to immersive or suggestive mental states.

In a previous study [58], we demonstrated a comparable pattern at the level of the autonomic nervous system, whereby baseline physiology predicted the magnitude of state-induced modulation. The present findings extend this model to the cerebral level, highlighting a physiological continuity between brain and body: individuals with particular baseline traits—perhaps akin to hypnotic susceptibility—may exhibit greater neurophysiological adaptability during NOC such as those induced by AICT.

Limits

This study presents compelling results underpinned by a rigorous methodological approach. However, several limitations warrant consideration. Firstly, the sample presents a notable gender imbalance, along with variability in age and levels of AICT expertise. These factors may introduce significant heterogeneity and limit the generalizability of the findings. Future research with larger and more demographically diverse samples will be essential to strengthen external validity. In addition, AICT expertise should be more explicitly controlled for or included as a covariate, as previous studies suggest it can impact both brain structure and function [108,109]. The use of EEG also presents methodological constraints. Traditional EEG systems require participants to remain physically still, which reduces the ecological validity of the trance induction and may have biased participants' phenomenological experience. Future work could address this limitation through the use of novel portable EEG systems (e.g., backpack EEG setups), allowing greater freedom of movement and more naturalistic exploration of these states. Finally, a more thorough exploration of the links between brain dynamics and phenomenological reports would be valuable. This could be achieved through more granular measures of subjective experience, and by applying longitudinal designs to assess whether repeated induction of trance states leads to cumulative neural changes and potential therapeutic benefits.

<u>Conclusion</u>

This work focuses on changes in electrical brain activity between a resting state and the AICT state. We demonstrated that the aperiodic component is the most effective in distinguishing between the two states, showing increased cortical excitability in frontal regions and the PCC during the AICT state. Additionally, a decrease in brain complexity in left posterior regions was also associated with the AICT state. Most of the changes in frontal and parietal regions appear to be strongly linked to significant individual variability already present at baseline, suggesting a key role for baseline brain states in these fluctuations.

AICT is a unique NOC distinguished by its specific characteristics, including the type of subjective experience it generates [7,44,45], the physiological changes [20,48,58] it induces and as demonstrated in this study, the associated changes in brain activity. We propose that the brain changes linked to this state support a profound and meaningful subjective experience with significant implications for both the individual's psychological and physiological functioning. We hope this work will inspire further research and contribute to a deeper understanding of the mechanisms that underlie the experiences associated with this state.

# Acknowledgments

We thank the participants for their involvement, as well as TranceScience Research Institute and Francis Taulelle.

# Research Transparency Statement

Conflict of interest disclosures: All authors declare no conflicts of interest, except that CS is the founder of the TranceScience Research Institute in Paris.

Founding disclosures: The study was supported by the University and University Hospital of Liege and its Algology Interdisciplinary Centre, the Belgian National Funds for Scientific Research (FRS-FNRS), the BIAL Foundation, the FNRS MIS project (F.4521.23),, the Belgium Foundation Against Cancer (Grants Number 2017064 and C/2020/1357), the Benoit Foundation (Bruxelles), the Fondation Leon Fredericq, and Wallonia as part of a program of the BioWin Health 619 Cluster framework, the FLAG-ERA JTC2021 project ModelDXConsciousness (Human Brain Project Partnering Project), the Mind Science Foundation, and the Horizon 2020 MSCA – Research and Innovation Staff Exchange DoC-Box project (HORIZON-MSCA-2022-SE-01-01; 101131344). VO was awarded the Francis Taulelle postdoctoral fellowship, funded by the TranceScience Research Institute. JA is postdoctoral fellow at the FWO (1265522N). KJ is supported by funding from the Canada Research Chairs program (950-232368) and a Discovery Grant from the Natural Sciences and Engineering Research Council of Canada (2021-03426), a Strategic Research Clusters Program (2023-RS6-309472) from the Fonds de recherche du Quebec – Nature et technologies. OG is research associate at FRS-FNRS.

Artificial intelligence disclosures: The text was revised and corrected using ChatGPT-4.

Materials: All EEG data are available upon reasonable request at avanhaudenhuyse@chuliege.be and ogosseries@uliege.be.

Script analysis: A GitHub repository consolidates all Python scripts used for machine learning, regression, and visualization (https://github.com/LIKACT/Phenomenology-Heart-rate-variability-in-Auto-Induced-Cognitive-Transe-AICT-).

## Authors contributions

Design of the study: O. Gosseries, A. Vanhaudenhuyse

Data collection: O. Gosseries, A. Vanhaudenhuyse, C. Martial, J. Annen

Data analysis: V. Oswald, K. Jerbi, O. Gosseries, A. Vanhaudenhuyse, H. Abdelhedi

Manuscript writing: V. Oswald

Reviewing paper: O. Gosseries, A. Vanhaudenhuyse, C. Martial, V. Oswald, K. Jerbi, C. Sombrun, J. Annen, H. Abdelhedi

*Supplementary Material*

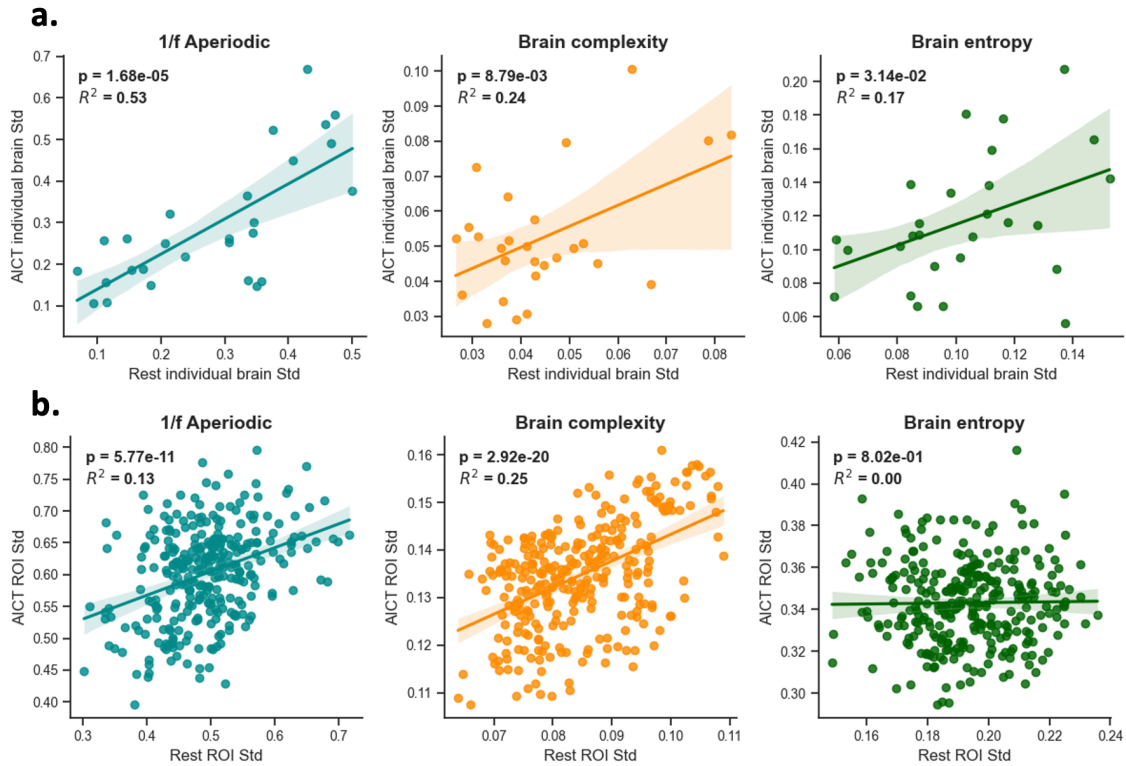

*Figure S1: Variability effect between Rest and AICT across individual and brain space.* *(a) correlation between individual brain variability (std) for each subject between Rest and AICT for 1/f aperiodic exponent, Lempel-Ziv complexity, and sample entropy. (b) correlation between group variability (std) for each ROI between Rest and AICT for 1/f aperiodic exponent, Lempel-Ziv complexity, and sample entropy.*